# Untethered Micro-Robots for Surface Sensing through Electric-Field Confined Motion

Huaizhi Li[a], Kyoungtae Kevin Park[a], and Donglei Emma Fan[a, b, c]

a. Materials Science and Engineering Program, Texas Materials Institute, The University of Texas at Austin, Austin, Texas, USA, 78712
b. Walker Department of Mechanical Engineering, The University of Texas at Austin, Austin, Texas, USA, 78712
c. Chandra Family Department of Mechanical Engineering, The University of Texas at Austin, Austin, Texas, USA, 78712

Correspondence and requests for materials should be addressed to the authors: Donglei Emma Fan (dfan@austin.utexas.edu).

**Abstract**

Surface characterization is essential for revealing the structural, chemical, and physical properties of materials. Yet high-resolution methods such as atomic force microscopy (AFM) require complex equipment and delicate skillsets, making them particularly challenging for applications involving soft and biological materials in liquids. Here, we propose and validate an innovative motion-enabled sensing scheme that uses untethered micromotors as robotic probes to interact with surfaces or objects, with their motion responses serving as sensing signals for characterization. This sensing concept is validated by employing 3D electrokinetic tweezers, which control micro/nanoparticles with up to 20 nm positioning precision in solution, to drive Au microsphere motors along designed scanning paths. When the motors encounter local chemical or structural variations, their locomotion changes, allowing motion itself to serve for the detection. This effort enables untethered motors, for the first time, to detect biomolecular patterns and lithographically defined microridge arrays in liquid environments. The work establishes robotic locomotion as a new sensing modality, opening a wireless, solution-compatible, and low-cost technical pathway to standard surface sensing.

## Introduction

Surface characterization plays a crucial role in understanding the chemical, physical, and topographical properties of materials, important to diverse fields, ranging from optoelectronics, biomedicine, mechanics, to life science. Traditional techniques such as atomic force microscopy (AFM) and scanning electron microscopy (SEM) provide high-resolution imaging and precise measurements but have grand challenges. These methods typically require controlled environmental conditions, complex instrumentation, and advanced skill sets, making them possible but challenging to use for soft biological samples and in liquid conditions. For AFM, enabled by a tip tethered to a one-end-fixed cantilever, the level of force is derived from the spring constant, *k*, of the cantilever, which typically ranges from 0.1 to 100 N/m. To reach cellular- and molecular-level forces in the pN range, it means that the deflection should be controlled to ~0.1 nm, requiring atomic-level precision. Thus, specialized AFM probes are made for biological samples with *k* values of ~0.01 N/m. This still requires gentle tip displacement of ~1 nm, highlighting the importance of equipment accuracy and user skill of AFM to detect molecular features. In contrast, untethered micro/nanorobots not only can be easily controlled in locomotion for targeted probing, but also offer an intrinsic force level in the pN range, corresponding to ~1 µm in size in water, when driven by widely used magnetic, electric, or optical fields. This force level, which is tunable by size, can readily match those of molecules and soft samples, such as gels. Moreover, untethered motors can offer multimodal sensing capability[1] through designed chemistry and structures, such as surface-enhanced Raman scattering (SERS) for biochemical detection[2]. Indeed, micromotors' abilities in controlled locomotion over large areas, wireless operation, and multiplexed sensing make them promising complementary tools for surface and fine structure and object characterization where traditional methods are much constrained[3].

Particularly, a micromotor's native motion in response to its environment has great potential for sensing applications. However, such a motion-based sensing mechanism has not been well explored. The most relevant studies are based on the equilibrium of a particle under two oppositely applied forces. One example is the use of optical tweezers for trapping a micro-object, whose trapping conditions[4,5] can be used to map flow profiles[6,7] and detect cellular[8] and molecular bonding forces[9]. Similarly, magnetic-tweezer-controlled motors have been applied to determine molecular forces and local viscoelasticity[10,11]. These measurements are based on balancing external force/torque with drag forces from molecules or fluids, both applied to the trapped particle. Intriguingly, a particle's Brownian motion itself, without applied external physical forces, can be used to infer the local temperature in solution[12]. In this case, thermal energy acting together with liquid viscous drag results in the observed Brownian motion, whose statistics determine the characteristic diffusivity in a solution. Nevertheless, Brownian motion is random and passive, and it does not permit active probing of targeted objects with sustained, controlled interactions.

In this work, we propose an innovative motion-based sensing method for revealing structural and chemical patterns and objects using 3D electrokinetic tweezers[13]. The tweezer is our recent invention with an embedded capability to counter both translational and rotational Brownian motion, enabling control of micro/nanoparticles with a positioning precision of 20 nm and an angular precision of 0.5° under a standard optical microscope. This capability could open a new pathway for controlling untethered microrobots along designed pathways to scan targeted locations or interesting entities on a surface. The micromotors' motion as a response of their interactions with the targets could naturally work as a sensing reporter for revealing regions with distinct structures, morphology, chemistry distribution, and electronic properties.

To validate this concept, we carry out systematic experimentation assisted by theoretical calculations. In experiments, we employ simple Au colloidal microspheres of 1.5 μm in size as robotic motors and control their movement forward and backward along designed one-dimensional (1D) pathways using a feedback-control algorithm. The motion versus time is obtained. Here, a key challenge in electrokinetic manipulation is the influence of electric double layers formed near electrodes, which nonlinearly change the electric-field strength in solution and are extremely difficult to study. To overcome this issue, we develop a switching-current compensation method to enable quantitative electric-field calculation with high accuracy. With this effort, we robustly detect both biomolecular patterns and lithographed microridge arrays on a surface, demonstrating the system's capability to distinguish both chemical and structural variations on surfaces using a motor's locomotion. Together, the reported motion-enabled sensing method expands the potential of micromotors and robots as untethered probes for non-invasive, wireless, and high-resolution surface characterization. The technique could potentially serve as a complementary technique to AFM while being much more affordable, easy to use, simple in equipment, and compatible with soft materials and solution environments.

## Working Principle

In a typical experiment, a micromotor, such as a Au sphere of 1.5 μm in diameter, is controlled by the electrokinetic traps to prob a targeted object. We choose Au microspheres, which are of high density and inert surface, to leverage their gravitational sedimentation effect to the substrate. This allows the Au spheres—acting as untethered motor probes—to remain close to a surface, ensuring consistent particle–surface interaction. Instead of the proportional-integral (PI) control algorithm used in our earlier work for precisely controlling particles' motion, we employ a bang-bang controller, which offers simple amplitude-constant electric outputs at 2 V ($V_A$) and -2 V (-$V_A$). During the operation, $V_A$ drives the propulsion. When the position of the motor exceeds the set point, - $V_A$ is applied to push the motor back to the set point; when the position of the motor falls below the set point, $V_A$ is applied to push it back. As a result, the motor oscillates around the set point. The position of the motor, $x$, is recorded at each video frame ($n$), along with the corresponding time, $t$. Therefore, the time and position at the $n$th frame are $t$[n] and $x$[n], respectively. The displacement ($d$) of the motor within frame $n$ is given by $d[n] = \Delta x[n] = x[n+1] - x[n]$. Using this displacement, the velocity $v$ can be calculated as $v = d/\Delta t$, where $\Delta t$ is the time interval between frames. The motor's mobility $\mu$ can be determined by $\mu = v/E$ at an applied electric field ($E$). For simplicity, we use $d$ within a single frame rather than $\mu$ to report the sensing signal at a given frame rate. This strategy enables us to directly compare $d$ with other quantities, such as imaging resolution and Brownian motion displacement.

As the mobility of the motor reflects the interactions between the motor and its environment, to further analyze the directional motion, we categorize the displacement as rightward ($d_{right}$) and leftward ($d_{left}$), under positive and negative voltages, respectively. In the absence of external interference, $d_{right}$ and $d_{left}$ have the same magnitude but opposite directions. Under this convention, both $d_{right}$ and $d_{left}$ are treated as positive quantities.

Under applied electric field, the velocity of electroosmotic flow is proportional to the substrate's zeta potential, expressed as $u_s = -\frac{\epsilon \zeta_s E}{\eta}$, where $\varepsilon$ is the permittivity, $\zeta_s$ is the substrate zeta potential, and $\eta$ is the fluid viscosity. If the substrate consists of different materials with varying zeta potentials, the mobility of the particle will differ depending on its location. Specifically, when the particle is over a region with higher or lower zeta potential, both $d_{right}$ and $d_{left}$ will be changed to be larger or smaller. In addition, particle responds to the same $E$-field under electrophoretic force, which is also proportional to the particle surface zeta potential, given by $u_p = \frac{\epsilon \zeta_p E}{\eta}$. As the particle is the same in $u_p$, by monitoring these directional displacements, which is the total of $u_s$+$u_p$, one can infer the local zeta potential of the substrate. Scanning the particle across different set points allows for the determining a zeta-potential profile of the surface.

If an external disturbing force $F_{ext}$ is present, it will act on the spherical particle and cause a movement. The force is balanced by the drag force, given by $F_{drag} = -6\pi\eta a v$, where $a$ is the radius of the particle. This results in a terminal velocity of $v_{ext} = \frac{F_{ext}}{6\pi\eta a}$. As a result, the total velocity of the particle becomes the sum of the velocity induced by the applied voltages, $v_V$, and the velocity due to external force, $v_{ext}$: $v_{total} = v_V + v_{ext}$. Importantly, $v_{ext}$ remains in the same direction regardless of the applied voltage, whereas $v_V$ changes sign when the polarity of the voltage is reversed. Therefore, we have:

$$d_{right} = \mu \frac{V}{d_{gap}} \Delta t + v_{ext} \Delta t, \quad (1)$$

$$d_{left} = -\left(-\mu \frac{V}{d_{gap}} \Delta t + v_{ext} \Delta t\right) = \mu \frac{V}{d_{gap}} \Delta t - v_{ext} \Delta t, \quad (2)$$

here $d_{gap}$ is the distance between the electrode pairs and the electric field is given by $E = V/d_{gap}$. From the above, we derive:

$$\frac{d_{right} - d_{left}}{2} = v_{ext} \Delta t, \quad (3)$$

$$\frac{d_{right} + d_{left}}{2} = \mu \frac{V}{d_{gap}} \Delta t. \quad (4)$$

By measuring $d_{right}$ and $d_{left}$, both the external velocity $v_{ext}$ and the particle mobility $\mu$ can be determined simultaneously. As a result, any physical effect that either exerts a force on the particle or alters its mobility can be probed at the same time. For example, the external Force $F_{ext}$ and zeta potential of the substrates $\zeta_s$ can be calculated using the following equations:

$$F_{ext} = \frac{3\pi\eta a (d_{right} - d_{left})}{\Delta t}, \quad (5)$$

$$\zeta_s = -\frac{\eta d_{gap} (d_{right} + d_{left})}{2\epsilon V \Delta t} + \zeta_p. \quad (6)$$

## Electric-Double-Layer Effect Analysis and Method

At the interface between the electrode and the solution, ions form an electric double layer (EDL) if a potential difference is presented. The EDL behaves like a capacitor, storing charge and producing a voltage drop denoted as $V_C$. When calculating the electric field strength $E$, the equation $E = V/d_{gap}$ is used, where $V$ is the effective voltage drop across the fluid. However, due to the presence of the EDL, not all of the applied voltage $V_A$ contributes to the electric field strength $E$ applied to a particle in solution. Instead, a portion of the voltage is consumed in charging the EDL. As illustrated in the equivalent circuit in **Figure 1**, the electrode-water system can be modeled as a series of resistors and capacitors. The EDL can be modeled by a capacitor $C_{EDL}$, the water between two electrodes can be modelled as a resistor $R_{water}$, and the electrochemical reactions occurring at the electrode-solution interface can be represented by a charge transfer resistor $R_{ct}$, where *ct* stands for charge transfer.

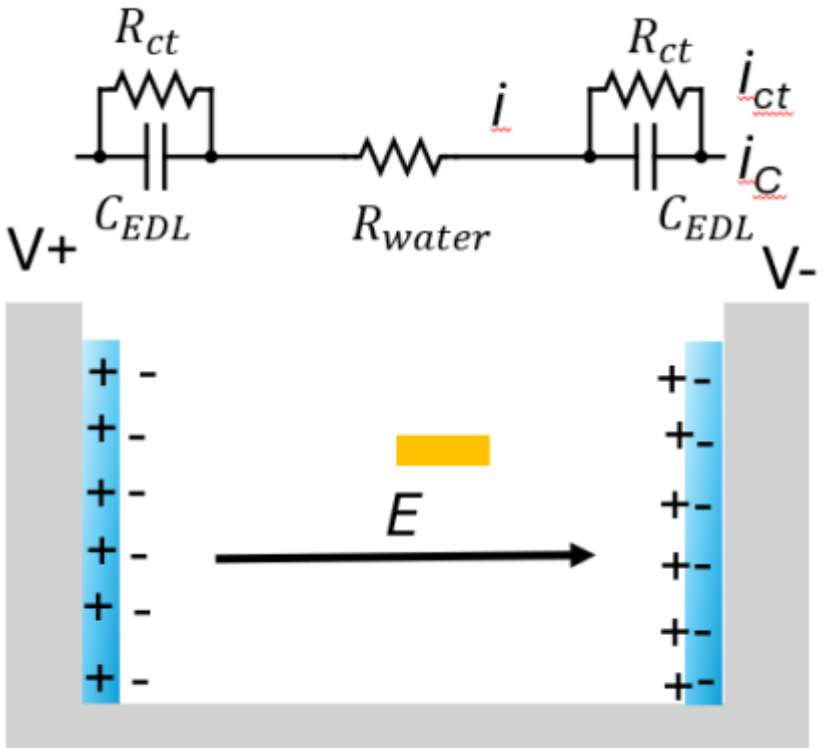

*Figure 1: A schematic of the electrodes system and the equivalent circuit.*

The applied voltage $V_A$ is distributed between the EDL capacitor $C_{EDL}$ and the solution resistance $R_{water}$. Since $R_{ct}$ and $C_{EDL}$ are in parallel, they share the same voltage. The total voltage stored across the capacitors is denoted as $V_C$, while the voltage across $R_{water}$ is denoted as $V_R$. This gives the relationship:

$$V_A = V_C + V_R. \tag{7}$$

Only $V_R$ contributes to the electric field strength, as given by $E = V_R/d_{gap}$. The EDL capacitance effect is clearly seen in **Figure 2**. When a constant voltage is applied between the electrode pairs, the velocity of the particle drops exponentially from -20 μm/s to near 0 μm/s instead of maintaining a steady velocity. This behavior arises because the EDL is charging over time, causing the voltage across water, as well as the electric field strength, to decrease exponentially. The red line in the figure is an exponential fit, showing excellent agreement between the experimental data and theoretical prediction.

Therefore, to accurately apply the technique discussed in the previous section, it is essential to determine the actual voltage drop across the fluid, $V_R$ , which is unknown, rather than relying solely on the known applied voltage, $V_A$.

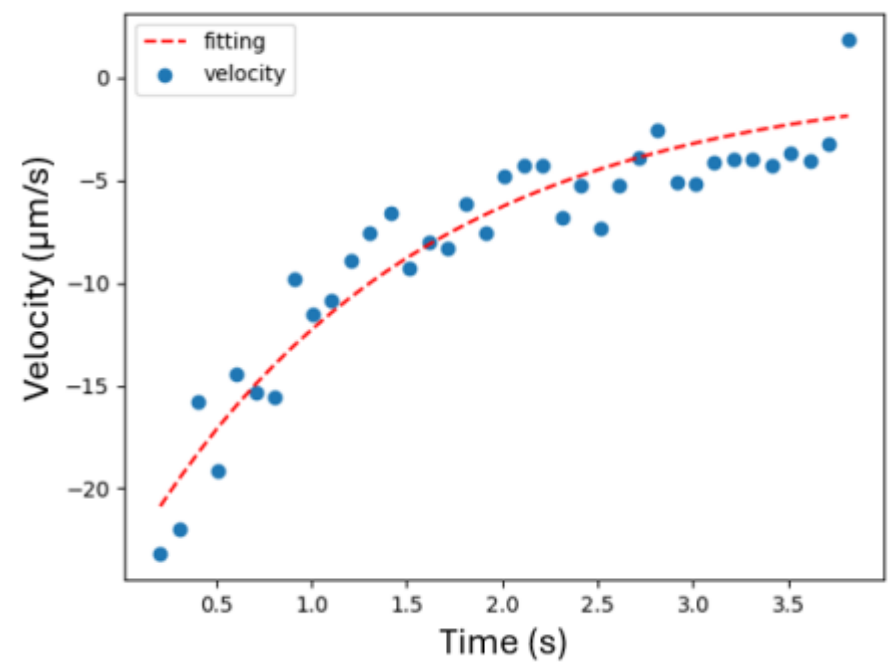

*Figure 2: The exponential decay of velocity of the particle due to the Electric Double Layer capacitance effect. Applied voltage: 0.75V.*

## Theoretical Analysis

Real-time measurement of the voltage drop across the EDL is challenging, as the capacitance of the EDL is a function of both the voltage and the ion concentration. In most of the literature[41], the applied voltage $V_A$ is used directly to calculate the electric field strength, while the capacitor effect is often ignored. In this section, a new method is introduced to determine both $V_C$ and $V_R$ by monitoring the current during the switching of the applied voltages.

Before switching, the applied voltage is $V_A$ and the capacitor is charged to $V_C$, and the voltage drop in water is:

$$V_{R,before} = V_A - V_C. \quad (8)$$

The corresponding current before switching is:

$$i_{before} = V_{R,before}/R_{water} = (V_A - V_C)/R_{water}. \tag{9}$$

After switching, the applied voltage becomes – $V_A$, but the switching is so rapid that both $V_C$ and $R_{water}$ remain unchanged. The new voltage drop in water becomes:

$$V_{R,after} = -V_A - V_C. \tag{10}$$

And the current immediately after switching is:

$$i_{after} = V_{R,after}/R_{water} = (-V_A - V_C)/R_{water}. \tag{11}$$

With simple derivation, we obtain:

$$i_{before} - i_{after} = 2V_A/R_{water}, \tag{12}$$

$$i_{before} + i_{after} = -2V_c/R_{water}, \tag{13}$$

$$\frac{i_{before}}{(i_{before}-i_{after})/2} = \frac{V_{R,before}}{V_A}, \tag{14}$$

$$\frac{i_{after}}{(i_{before}-i_{after})/2} = \frac{V_{R,after}}{V_A}. \tag{15}$$

By measuring the current immediately before and after each voltage switch, and knowing $V_A$, we can calculate $R_{water}$, $V_C$ and $V_R$. Notably, $R_{water}$ may vary over time due to ions generated from electrochemical reactions. Thanks to the nature of the bang-bang controller, where the voltage continuously switches between $V_A$ and -$V_A$, these values can be calculated at each switching event. The values between switching events can be estimated by interpolation, providing continuous monitoring of $R_{water}$, $V_C$ and $V_R$ throughout the experiment.

## Experimental Validation

The control algorithm determines when to switch the applied voltage by comparing the current position of the particle to the set point. As a result, the duration of each applied voltage pulse can vary before switching to the opposite voltage. **Figure 3** shows the measured current as a function of time when a particle is trapped by the bang-bang controller. The time intervals between voltage switches vary, as the voltage may persist for more than one frame depending on the particle's motion. When the voltage is switched, the measured current also flips sign. Between switching events, the current decays exponentially because of the charging of the EDL. The values of $i_{before}$ and $i_{after}$ can be determined from the plot and used to calculate $V_R$ and $V_C$ using Equations (10 -13).

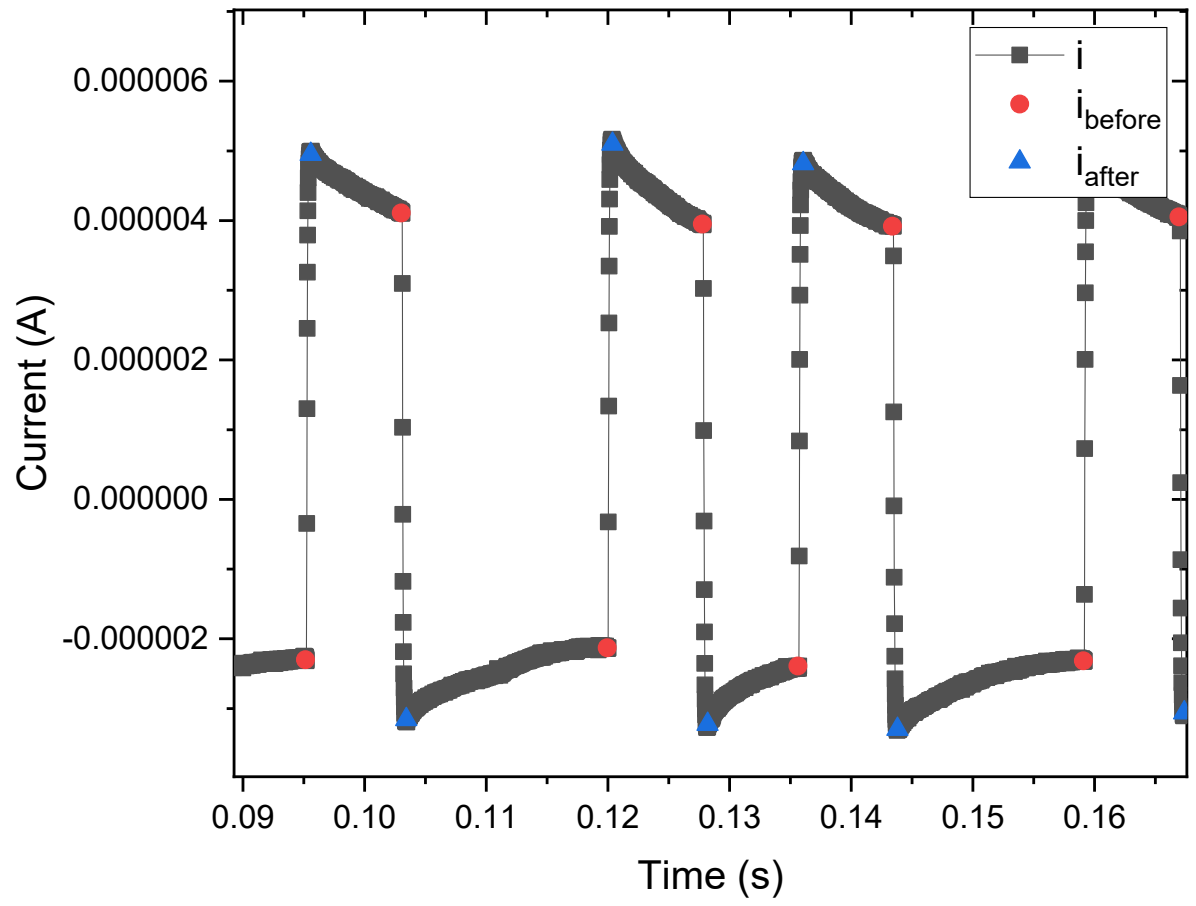

*Figure 3: Plot of measured current, $i_{before}$ and $i_{after}$ when a particle is trapped by the bang-bang controller.*

To perform a scan rather than a single-point measurement, the set point is moved sequentially according to a predefined scan pattern. At each set point, the particle is controlled for a specified duration to collect data. In a line scan, a net average voltage is needed to move the particle from one side to the other. This net voltage gradually charges the EDL, increasing the EDL voltage to $V_C$ and reducing the effective voltage drop across the fluid $V_R$, as shown by $V_R = V_A - V_C$. As a result, the particle's displacement in the direction of the scan will decrease. On the other hand, when the feedback voltage switches sign, the EDL discharges and the stored voltage superimposes with the applied voltage $-V_A$, leading to a larger magnitude of $V_R$, given by $V_R = -V_A - V_C$. As a result, the displacement opposite to the scanning direction will increase. This effect is clearly seen in **Figure 4a-b** during rightward and leftward scans on a blank substrate, where $d_{right}$ is smaller than $d_{left}$ during a rightward scan, and larger than $d_{left}$ during a leftward scanning.

To correct for this capacitive effect, $V_R$ is calculated using the switching current method. Since the raw displacement is proportional to $V_R$ instead of $V_A$, the observed discrepancy between $d_{right}$ and $d_{left}$ arises from variations in $V_R$ during the scan. To account for this, the raw displacement is normalized by dividing it by the ratio $V_R/V_A$. The corrected displacement corresponds to the normalized displacement that would be measured if $V_R$ equals $V_A$. As shown in Figure **4c-d**, $d_{right}$ and $d_{left}$ align closely after the correction. As a result, the mobility ($\mu$) can be calculated using the corrected displacement ($d_{right.c}$ and $d_{left.c}$), applied voltage ($V_R$), frame rate, and electrode gap distance, and is proportional to the corrected displacement. Since the corrected displacements overlap, this indicates that the mobility is equal in both the rightward and leftward directions, which is expected for a blank substrate, validating the method discussed above.

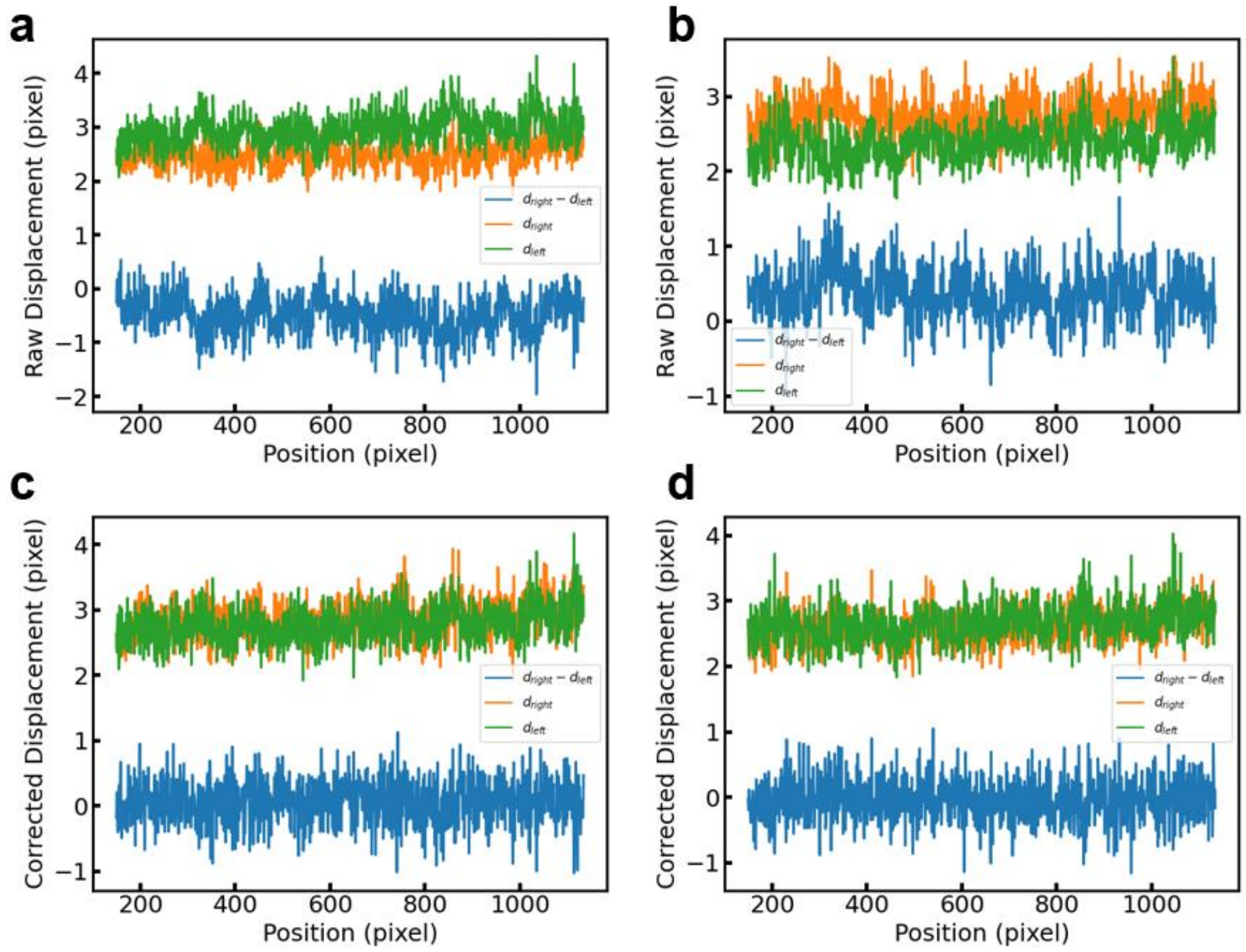

*Figure 4: (a-b) Raw displacement for (a) rightward and (b) leftward scanning, and (c-d) The corrected displacement for (c) rightward and (d) leftward scanning on a substrate uniformly coated with Bovine Serum Albumin. Frame rate: 169 FPS; applied voltage: 3V; electrode gap: 500 µm.*

## Micromotor Scanning Robots for Motion-Enabled Biomolecule Pattern Detection

To create molecular patterns on a substrate, we employed the microcontact printing method.[83,84] The experiment begins with patterning photoresist on a substrate followed by its passivation in a desiccator under vacuum along with 50 µl of Perfluorooctyltrichlorosilane (Alfa Aesar). Next, Polydimethylsiloxane (PDMS) precursor solution is introduced and cured on the substrate. This step transfers the substrate pattern to the PDMS stamp. Before printing Bovine Serum Albumin (BSA) to the microchip, the PDMS stamp side is treated with UV Ozone for 30 mins to generate silanol groups, followed by immersion in 100 µg/ml BSA (Sigma-Aldrich) in Phosphate-Buffered Saline (PBS) buffer for 45 minutes. After rinsing with PBS (1×) and deionized water (3×), and airflow drying, the stamp is brought into contact with the microelectrode for 5-10 seconds, transferring the BSA molecular pattern to the testing chip.

Next, a Au micromotor solution is pipetted into the PDMS well. The system utilizes a bang-bang controller, which applies either a positive voltage $V_A$ or a negative voltage $-V_A$, to keep the particle to the set point. A digital multimeter (Keithley DMM6500) is connected to measure the currents passing through the system. The position of the particle, the applied voltage, and the corresponding current are recorded, and the results at each point is calculated correspondingly. The system supports both line scanning and map scanning by dynamically updating the set point along a line or across a region of interest. At each set point, data are collected over a defined sampling period, and calculated. In addition to processing data according to the particle's set point, the data can also be processed based on the particle's actual position, permitting post-processing and more precise spatial mapping.

## Imaging Analysis Algorithm

We use OpenCV to analyze instant positions of a micromotor from its image. As the motor is spherical, orientation is not important, and we only need to determine the positional information. Therefore "cv::moments" is used, which calculates the center of mass by averaging the positions of all pixels in the particle's contour. As shown in **Figure 5**, this approach yields continuous positional values and high precision within 2 pixels in variation, which could be due to the minor stage drift over time, combined with vibration noise of the microscope body, camera, chip, and camera's image noise. The imaging algorithm selection is highly important for obtaining accurate displacement measurements and mobility analysis.

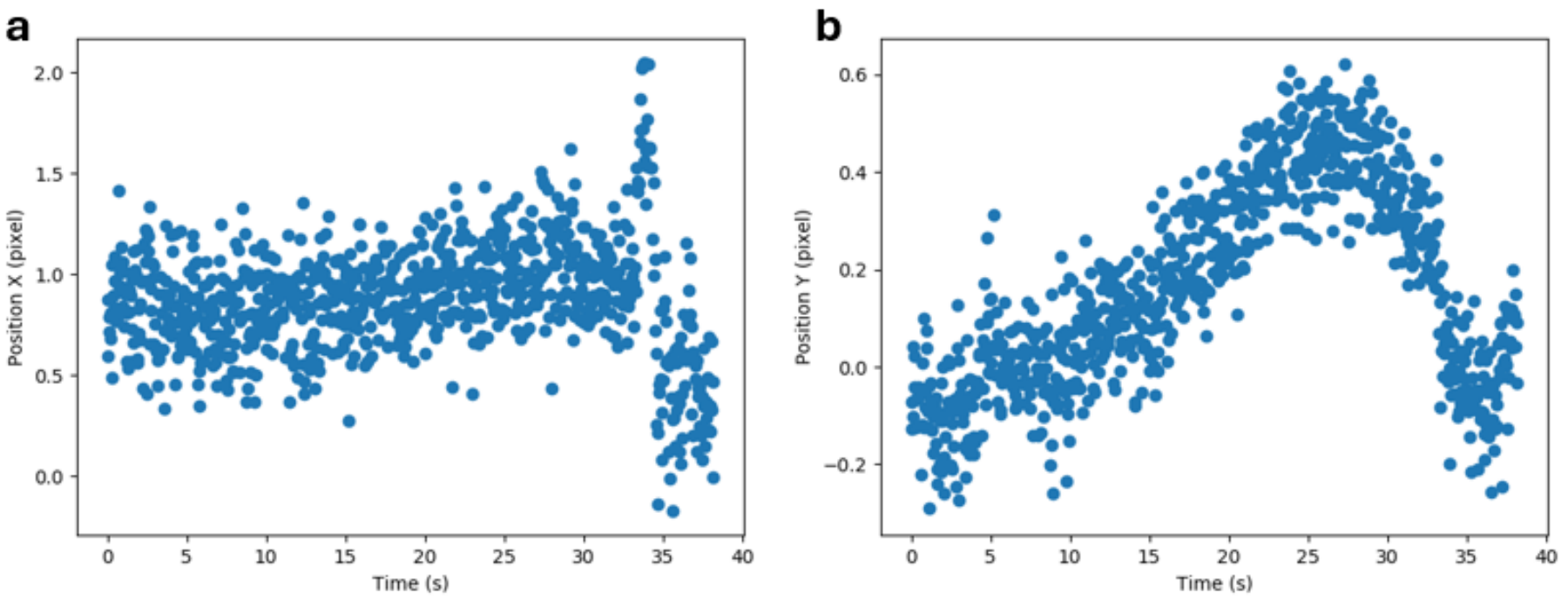

*Figure 5: Imaging precision of the "cv::moments" method. Position versus time for a stationary 6-µm nanowire in deionized water in (a) X and (b) Y directions versus time.*

## *Molecular Pattern Detection*

We scan the Au micromotor across an array of patterned BSA strips of 20 µm in width and 20 µm in gap. The strips are aligned and extended along the Y direction. A particle is scanned along the X direction, allowing it to travel across a region of alternating glass–BSA strips. The measured mobility on glass is consistently higher than that on BSA (**Figure 6**), showing the corresponding alternating strip patterns of BSA on glass. The results validate the concept of using motion of a controlled micromotor to detection variation of surface chemistry with high spatial resolution.

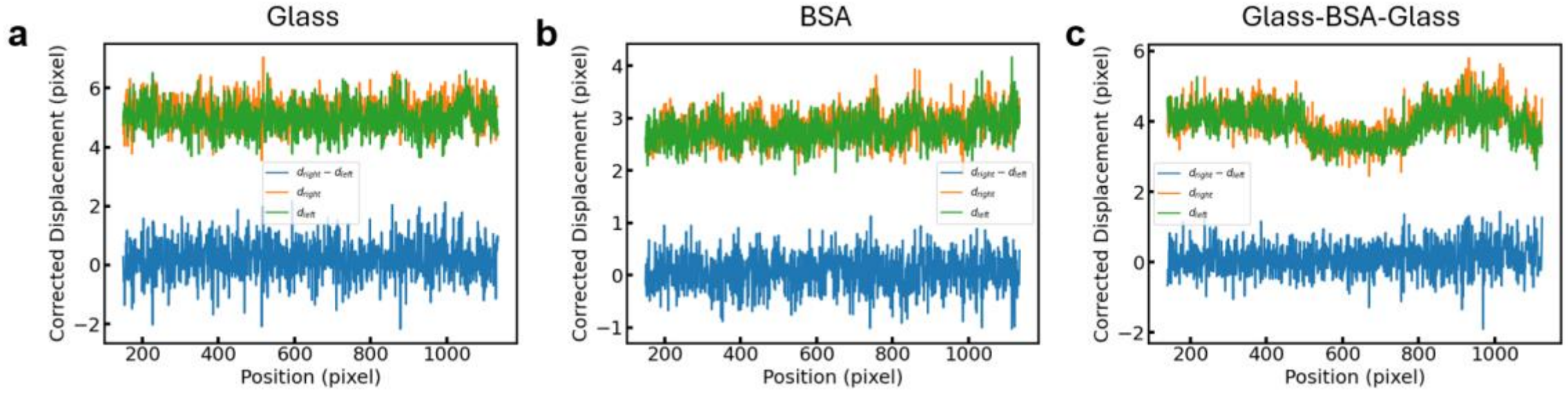

Figure 6: Displacement versus position in a line scan on (a) glass, (b) BSA, and (c) glass-BSA-glass-BSA alternating patterns. Scale bar: 1 pixel = 0.08 µm.

We further test the scanning along X directions at different Y coordination. Here the pattern varies only along the X direction and remains same in chemistry along Y. The different scans performed along Y at 268 pixels and 753 pixels, respectively, show that same level of mobilities are determined (**Figure 7**) further validating the technique's ability to characterize surface properties. The same results are obtained regardless of rightward or leftward scans.

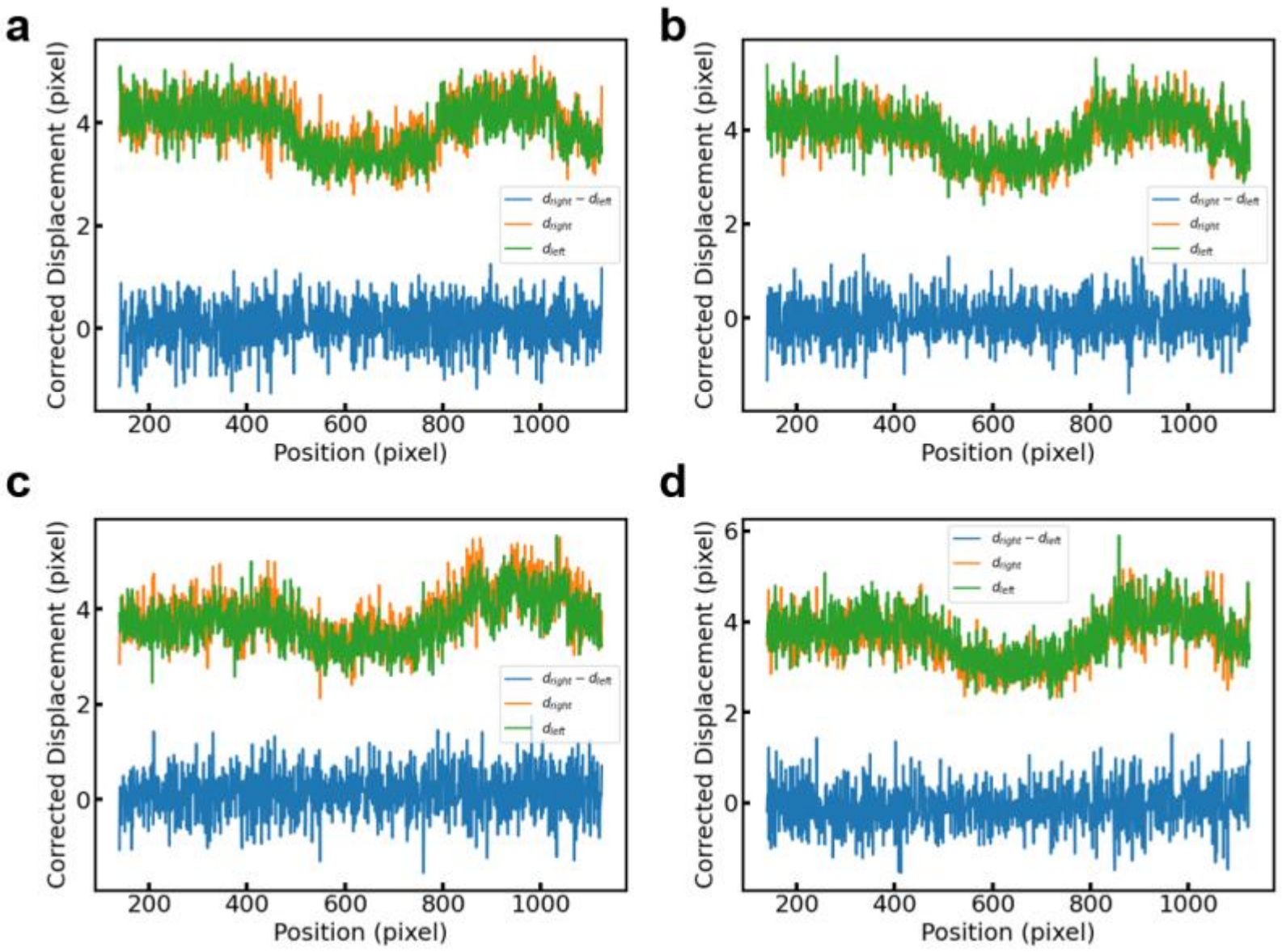

*Figure 7: (a-b) Electrical-double-layer (EDL) effect corrected displacement versus position at Y = 268 pixel, where (a) is the rightward scan and (b) is the leftward scan. (c-d) EDL-effect Corrected displacement versus position at Y = 753 pixel, where (c) is the rightward scan and (d) is the leftward scan.*

The capability in detecting and characterizing the dimensions of patterned biomolecules is further confirmed for a BSA pattern made with periodic of 2 µm in the through-pattern scans (Figure 8 a-c) and between pattern scans (Figure 8 d-f). Here the scanning is pixel based at 42 nm/px corresponding to ~6.3 um. We notice that the mobility of the micromotors show a gradual transition between the glass surface and BSA patterns, which could be due to the diffusion of BSA molecules as well as the size of the micromotor probe of 1.5 µm in diameter, which is on the same scale of the patterns.

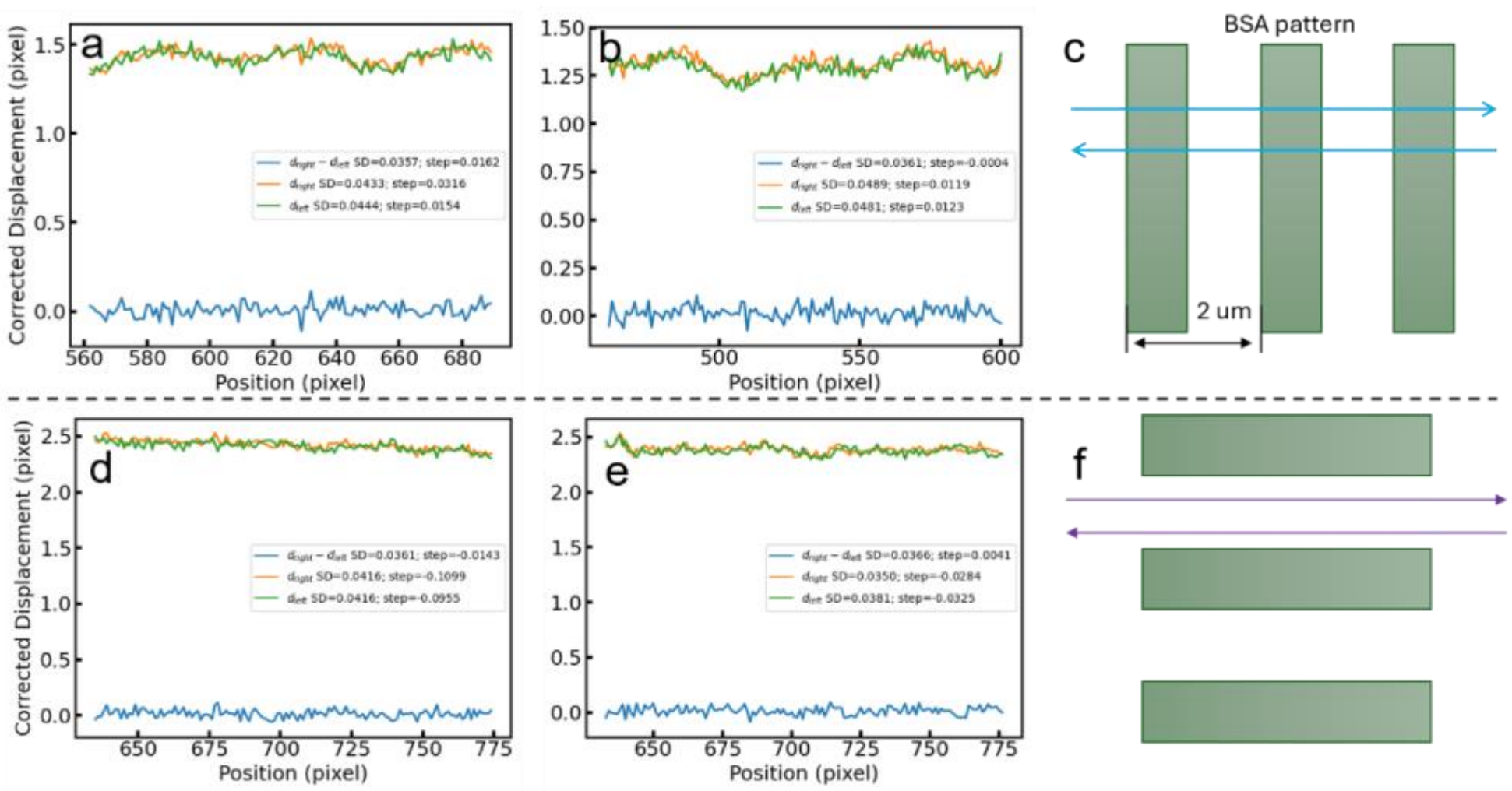

Figure 8. Scanning through patterned BSA molecules on the chip with periodicity of 2 μm. The through-pattern (a) forward and (b) backward scans correspond to the (c) patterned molecules. The between-pattern (d) forward and (e) backward scans following paths in (f) further support the robustness of the motion-based characterization.

## Profiling Surface Structures

When a structured object presents on the surface, collisions between the scanning motor and the object can instantly change the motor's motion. For example, if the object is located to the right side of the particle, it can exert a 'kick' that bounces the particle toward the left. As a result, $d_{left}$ can be larger than $d_{right}$, making $d_{right} - d_{left}$ negative. Vice and verse when the object is on the left side of the particle.

To test this understanding, we pattern periodic photoresist strips (4.5 μm in width; 7.5 μm in gap) on a quadruple electrode (**Figure 9a**). The thickness is controlled to be ~1 μ m. A 1.5-μm Au sphere is trapped and scanned across the strips towards the right in X direction at 4px/s through 240 targeted locations, across 960 pixels in total. At each pixel, $d_{right}$ and $d_{left}$ is calculated and plotted versus position (**Figure 9b;** $d_{right}$ orange, $d_{left}$ green) As expected, when the particle is near the left edge of the strip, ($d_{right} - d_{left}$) (**Figure 9b**, blue curve) becomes negative, indicating resistance from the pattern on the right. Conversely, when the particle is near the right edge of the strip ($d_{right} - d_{left}$) become positive. The effect is robust, repeated during the scanning of the patterning. The signal pattern well correlates to the spatial pattern, indicating success in profiling the structural features of the surface. During the measurement of the mobility of the micromotor along the X-direction, we also monitor its mobility along Y-direction. Also shown in **Figure 9c**, the transport also shows periodic with well overlapped $d_{right}$ and $d_{left}$. As aforediscussed, ($d_{right}$+ $d_{left}$) reflects the surface chemistry differences during scanning. The measured signals of ($d_{right}$+ $d_{left}$) is periodic, well corroborating with the respective positions of photoresist strips and glass-surfaced gaps. The results further reveal the distinct chemical nature of photoresist and glass. Because $d_{right}$ and $d_{left}$ in the Y direction essentially overlaps, resulting in a flat line of ($d_{right}$- $d_{left}$) in blue, it indicates that the structural difference in Y direction is minimal, agreeing with the longitudinal feature of the strip patterns. Finally, the micromotor is propelled to scan along the Y direction on the glass area between strips. The results show clear flat-featured signals in both structure and chemistry, further validating the method (**Figure 10**). Overall, the scanning measurement and analysis not only unravel the structures of the pattern but also its chemistry variation, demonstrating the effectiveness and consistence of the proposed sensing and analysis method.

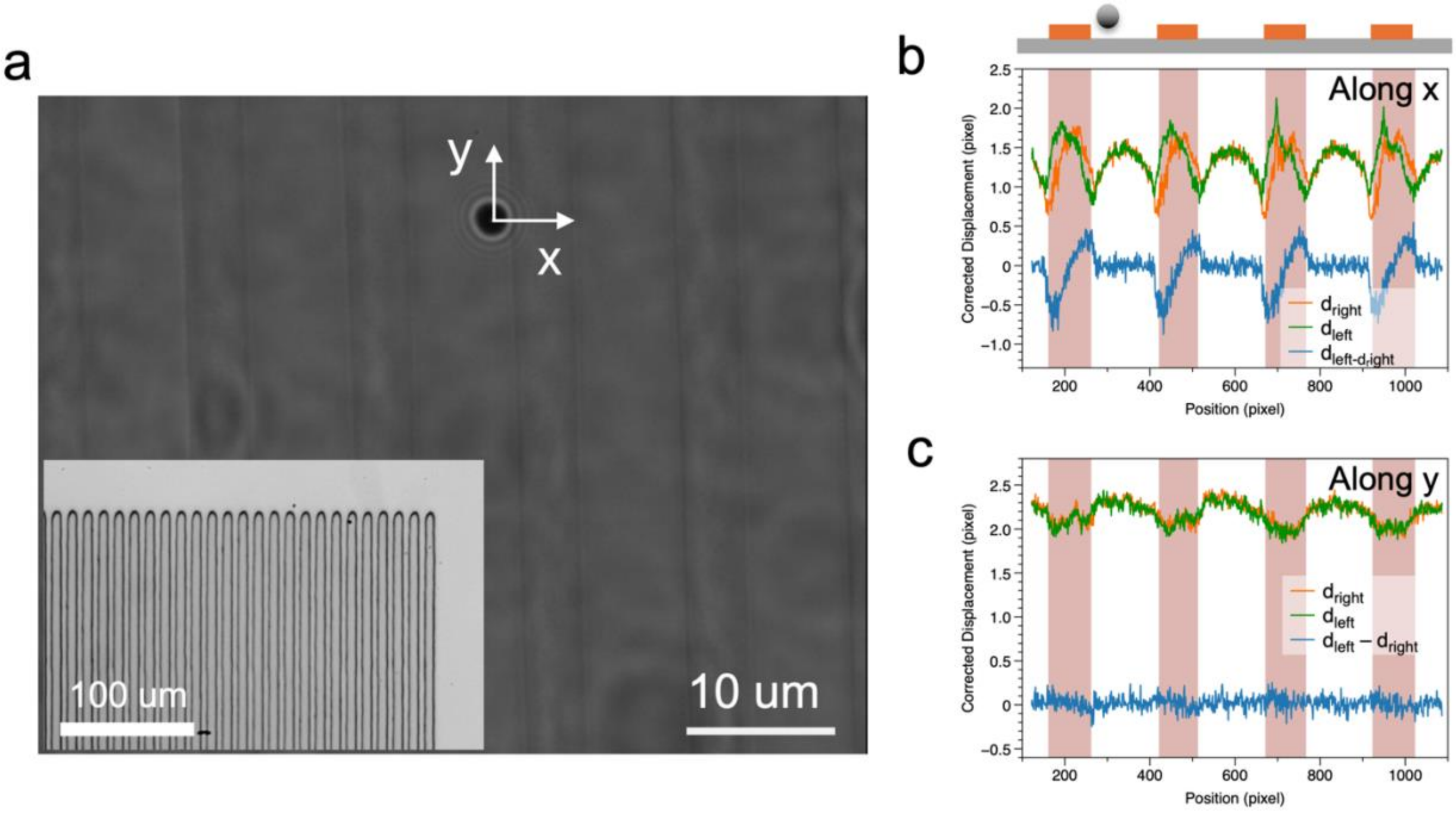

Figure 9: (a) Optical microscopy image of a gold colloidal micromotor on the patterned photoresist strips. The inset shows an optical microscopy image of the patterned photoresist strip array. The scanning is conducted from right to left in the direction of X-axis. (b) In X-direciton, ($d_{right}$- $d_{left}$) correlates with the structural features and (c) ($d_{right}$+ $d_{left}$) in the Y axis reveals the chemical differences between photoresist strips and glass surface. Here $d_{right}$ and $d_{left}$ in the Y direction essentially overlaps, resulting in a flat line of ($d_{right}$- $d_{left}$) in blue, which agrees that during the scan, the structural difference in Y direction is minimal.

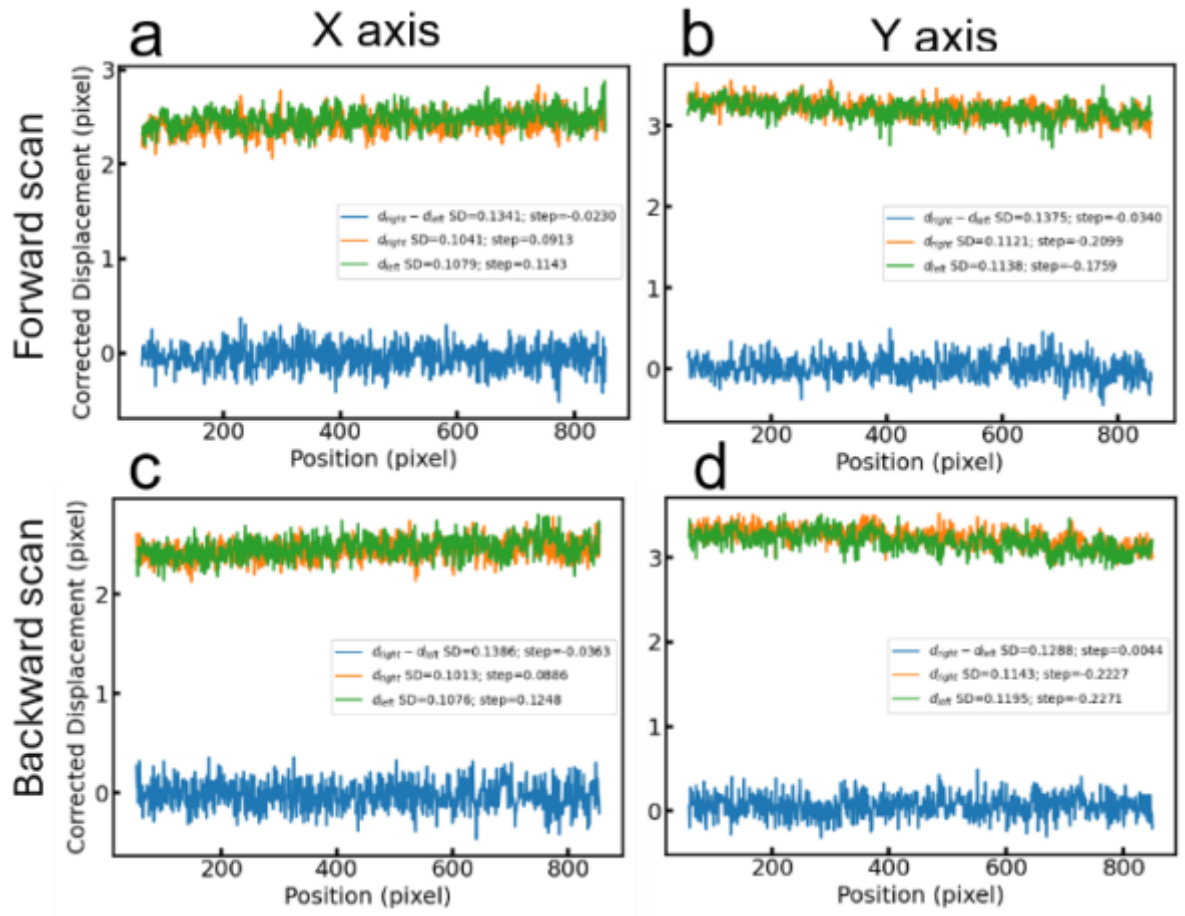

Figure 10. The micromotor is scanned along the Y axis on the glass area between the photoresist strip patterns, both +Yand -Y directional scans show flat features for both structure *(*$d_{right}$- $d_{left}$*)* and chemistry *(*$d_{right}$+ $d_{left}$*)* of the scanned area. The results align with the characteristics of the flat uniform glass substrate.

## 5.6 Discussions

This work proposed and successfully demonstrated that electrokinetically trapped, untethered colloidal micromotors can serve as sensitive scanning probes for surface characterization. By measuring directional mobilities under bang-bang control, the system is able to detect variations in both surface chemistry and structure. Changes in surface chemistry, such as differences in zeta potential, are reflected in the total magnitude of two directional scanning mobilities along the same axis. In contrast, topographical features are revealed by the difference between the two directional scanning mobilities, which capture the mechanical interactions between the particle and the surface. The switching-current-measurement method is original, enabling accurate accounting for electric double layer (EDL) effects, which significantly improves the reliability of the mobility calculations and sensing performance. The demonstrated capability in probing the EDL-associated parameters may also benefit other electrokinetic applications.

Here, we particularly emphasize the choice of the bang-bang control scheme, offering alternating application of positive and negative voltages, serves important purposes: (1) it maintains confinement of the particle near the set point by counteracting Brownian motion; (2) it generates switching currents that allow for real-time compensation of EDL capacitive effects; and (3) it readily enables the separation of forward and backward displacements along a direction, enabling simultaneous sensing of both the motor's mobility on a surface and external instant forces to the motor.

Most state-of-the-art micro/nanoparticle manipulation techniques lack nanometer control capability over particle motion. In contrast, the method described here enables active, controlled scanning of the substrate with high spatial resolution of < 100 nm. The ability to trap and move a particle along predefined paths while continuously collecting motion signals allows for robust scanning and mapping capability of patterns and objects on a surface. With its wireless and non-invasive nature, high control precision, and compatibility with fluid environments, the electrical driven scanning micromotors provide an original, versatile, low-cost platform for applications in surface and structure sensing, soft material characterization, and microfluidics where traditional cantilever-based methods or other untethered microrobots are often inadequate.